\documentclass[twocolumn]{aastex631}

\usepackage{lipsum}  % For sample text
\usepackage{graphicx}  % For setting text width in table
\usepackage{array}  % For column width control

\begin{document}

\title{Applying the Principles of 
Universal Design to Make Astronomy More Accessible}

\author[0000-0001-8481-2660]{A. C. N. Quirk}
\affiliation{Department of Astronomy, Columbia University, New York, NY, USA}

\author[0000-0002-7231-7328]{T. S. Rice}
\affiliation{American Astronomical Society}
\affiliation{Department of Physics, George Washington University, Washington, DC, USA}

\begin{abstract}
Universal Design (UD), an approach to accessibility that was first conceptualized in architecture to make buildings physically accessible, has since been applied to curriculum design to make classrooms accessible for a larger range of learning needs. In this paper, we illustrate how the concepts of UD are relevant outside of architecture and the creation of curricula by highlighting examples of norms that exist in the field of astronomy that create barriers for disabled folks. we discuss ways the foundations of UD can be applied more generally to department culture, conferences, outreach events, and academia as a whole to make STEM fields more inclusive. In order to implement UD in these sectors, one must create multiple pathways or options for folks to engage with and show their success in astronomy. While UD is critical for disabled folks, it can easily be expanded to include the promotion of people whose backgrounds and/or identities are currently underrepresented or under-supported in STEM. Lastly, we introduce guiding questions and tools for departments and institutions to evaluate the accessibility of their activities and traditions to disabled individuals. In summary, we aim to show the importance of increased accessibility and provide some strategies to make STEM more inclusive to disabled people by using the mindset and principles of UD.

\end{abstract}


\section{Introduction} \label{sec:intro}
\subsection{The State of Disability Inclusion in Astronomy}

According to a 2018 study of US members of the American Astronomical Society (AAS), 18\% of US astronomers considered themselves to have a disability \citep{pold2019}, which is slightly lower than the percentage of Americans that are disabled, 28.7\% \citep{CDC2024}. (The global percentage is more similar at 18\% \citep{WHO2024}). This number by the AAS is likely an underestimate because of existing stigma. Since there is a substantial fraction of astronomers who are disabled, efforts have been made to accommodate disability. To promote the inclusion of and opportunities for disabled astronomers across all career fronts, the Working Group on Accessibility and Disability was created as part of the AAS \citep{WGAD} in 2016. Disability was an important topic at both Inclusive Astronomy 1 and 2 conferences \citep{inclusive1, inclusive2}. There have also been conferences centered on disability in STEM or astronomy (e.g., \citealt{sciaccess} and the ArXiv Accessibility Forum \citep{arxiv}), highlighting the interest in and need to discuss the experiences of disabled people in STEM. However, despite the attention disability has gotten in astronomy in the past decade, there has been a lack of systemic change to make the field more inclusive. The current system requires individuals to get an accommodation from their institution, which almost always requires official medical documentation, disclosure, scrambling, and temporary changes. It also puts the burden on the individual to get the support they require. Universal Design (UD) is an approach to disability accessibility that can offer an improvement to the status quo described above. 

\subsection{Universal Design} \label{sec:architecture}
UD is a concept that originated from the field of architecture and was first coined by Ronald Mace \citep{story1998universal}. It is the mindset that physical spaces should be built to be accessible instead of later retrofitted to be accessible. This places accessibility at the very beginning of planning. It also recognizes that later retrofitting, while important if needed, is usually less effective, expensive, invasive, and can conflict with the initial design purpose of a space \citep{story1998universal}.

\par An example of the use of Universal Design is a curb cut. A curb cut is the dip in the sidewalk where crosswalks meet the curb. This dip allows wheelchair users, or someone using a mobility device, to cross the street safely and easily. Without the curb cut, they do not have safe access to going to and from the sidewalk/street. The curb can be retrofitted to have a ramp, but that requires additional construction and maintenance. While the curb cut is critical for folks with physical disabilities, its benefits are not limited to them. Anyone, disabled or not, can use a curb cut. These dips allow parents pushing strollers, delivery people pushing dollies, etc, safe and easy transportation. In this sense, UD can benefit non-disabled people as well. 

\subsection{Universal Design for Learning}
The mindset of UD has become popular in education, and has been applied to create Universal Design for Learning (UDL; \cite[e.g.,][]{rose2000universal, rose2001universal, hitchcock2005equal}), which was first created in the 1980s by the Center for Applied Special Technologies \citep{cast}. Like in architecture, UDL involves creating lessons, curricula, etc, with accessibility in mind from the beginning. This requires an instructor to make clear and straightforward paths for students to work with information and to demonstrate their success. The principles of UDL involve creating multiple pathways along three axes: representation, action and expression, and engagement \citep[e.g.,][]{cast, ralabate2011universal}. For example: an ideal instructor would create multiple assessment options, such as an exam or a project, so that students can choose an option that is most friendly to their disability, learning preferences, and motivations.

\par Just like UD in Architecture, UDL is critical for disabled students but will benefit all students. There is significant research on the benefits of implementing UDL as can be found in analyses like \cite{capp2017effectiveness, spooner2007effects}.

In this paper, organized as follows, we describe how UD can make astronomy more accessible not only for those with disabilities but for all individuals. In Section \ref{sec:astro}, we discuss how UD can be applied to the field of astronomy and implemented as an anti-racist tool. In Section \ref{sec:pathways}, we give examples of how the field can become more flexible and expand ways astronomers are represented, engage with work, and show success, and in Section \ref{sec:appendex}, we highlight some beginning tools one can use to increase the accessibility of their department or institution. The contents of this paper rely on my perspective, as developed by my lived experiences and interactions. we will highlight how the range of disability is vast, so it is inevitable that we have not included all of the angles we should consider in order to make the field of astronomy accessible. 

\section{UD Applied to the Field of Astronomy} \label{sec:astro}
The Astro2020: Panel on State of the Profession and Societal Impacts highlights the importance of multimodal expertise, in which there are "multiple ways of prioritizing, assessing, and evaluating knowledge, including the science and research objectives of the field" in order to create a more inclusive field \citep{decadel}. we believe that we can apply the mindset of UD and the principles of UDL more broadly to the field of astronomy in order to achieve this. UD can be applied to university departments, research institutions, outreach events and organizations, and the structure of academia. The principles of UD state that one should create multiple pathways (or options) for how folks participate, in this case, in astronomy. We can make plans, policies, etc, with accessibility in mind from the very beginning. We can also make those plans, policies, etc, known in advance and explain the reasoning behind them, since creating clear pathways to success and participation is an important part of UD. This allows people to choose a way that is most accessible for their specific needs and will help minimize the number of accommodations needed. When making multiple pathways, it is critical that all pathways are seen as equally valid, instead of as alternatives for disabled people. In Section \ref{sec:pathways}, we discuss three axes to focus on when creating multiple options. 

\par However, it is important to remember that the range of disabilities and individual needs is quite wide and varied, and sometimes needs can be at odds with one another. Because of this, one can rarely design a policy to be perfectly accessible to everyone. Accommodations are still going to be necessary, so it is critical to create a safe environment where folks feel comfortable to identify a barrier they have encountered. In order to foster such an environment, one should be flexible with policies, not further stigmas, mirror someone's language when describing their disability, make it clear you want everyone to succeed regardless of disability, and ask for and incorporate feedback. See the Appendix for more details on creating a department that is disability friendly. 

\subsection{Intersectionality}
An added bonus of UD is that it can benefit folks who aren't disabled as well (similarly to the example of the curb cut in Section \ref{sec:architecture}). While it is critical for disabled folks, it is a tool that can support other identities and backgrounds that have been or are currently underrepresented and under-supported in STEM. When creating multiple pathways for people to participate in astronomy, one can bring in the idea of intersectionality. Intersectionality was coined by Kimberlé Crenshaw and is a framework to explore how society interprets and acts on an individual's multiple identities, which results in a unique level of privilege and/or discrimination \citep{kimberly1989demarginalizing}. Folks are not just disabled or non-disabled. Everyone has other identities, such as race, class, gender, sexuality, age, etc, that can be further supported by UD.

\subsubsection{UD as an Anti-Racist Tool}
Intersectionality tells us not all disabled individuals are treated in the same manner. Studies show that Black and Latine children with developmental delays are 78\% less likely to receive early support \citep{feinberg2011impact}, and Black disabled students are twice as likely to be disciplined than white disabled students \citep{NYT}. Furthermore, 30\%-50\% of Black people killed by the police have a disability \citep{perry2016ruderman, mccauley2017cumulative}, indicating that systemic racism leads not only to less support but also to more violence towards Black disabled individuals. Racism also can cause poor mental health in individuals experiencing it \citep{chakraborty2002does, burke2018racism}, and majority-Black communities have less access to healthcare \citep{yearby2018racial}, so someone who is BIPOC has the chance to become disabled because of systemic racism. 

\par Getting an accommodation involves the healthcare system and the HR system, which requires folks to trust these systems that are rooted in systemic oppression \citep{yearby2018racial}. If an individual has been harmed or invalidated by these systems in the past, they might be less likely to use them going forward. Thus, the current accommodation system is likely to benefit white disabled people more than BIPOC disabled people, since white disabled people are more likely to have been supported by educational and health interventions. In this sense, implementing UD, which minimizes the accommodation system, is one way to practice anti-racism. 

\section{Creating Multiple Pathways}\label{sec:pathways}
When creating multiple pathways, we propose focusing on the three axes of UDL: representation and motivation, engagement, and action and expression \citep{cast, ralabate2011universal}. In the next sections, we will describe each axis and highlight some of the norms that exist in astronomy. We will discuss how these norms harm disabled folks and how they also might negatively affect non-disabled people who have other impacted identities.

\subsection{Representation and Motivation} 
\subsubsection{Astronomers}
The most basic question with regards to representation is how astronomers are represented in academia, media, outreach, etc. The norm, like in the rest of society, is to hide disability or the impact of one's disability. Because there is still a societal stigma around being disabled, members of the astronomy field might not disclose their disability. This is of course a feedback loop: the more one hides their disability, the more it seems as if there are no disabled astronomers. An easy solution is to highlight disabled astronomers and their work. There have been, are, and will be astronomers with disabilities. This is a great opportunity to bring in intersectionality as well and be sure that it's not only white disabled astronomers that are highlighted in teaching, talks, and outreach programs. 

\par It is a personal decision for someone to disclose if they are disabled or not. Some, if they are visibly disabled (i.e. use mobility devices), might not have a choice about disclosing their disability but instead might try to hide or minimize the impact of it. Others with invisible disabilities (i.e. chronic pain or mental illness) might choose to keep their disability private, as is their right to do. While it is up to the individual, if someone chooses not to disclose a disability at their workplace or to their peers, it is likely because they feel they will not be fully supported. Thus, it's important for departments to actively celebrate and support disability so that more disabled astronomers are comfortable describing themselves as disabled and asking for support. The more disabled folks can safely and comfortably label themselves as disabled in STEM fields, the less stigma there will be. 

\subsubsection{Data}
\par While how we represent the people who do astronomy is important, we must also think about how data is represented. Almost all astronomy data products require one to be sighted, from plots to processed images. While technology like text-to-speech programs and screenreaders allow blind people to access written work, images are not fully accessible. Images should be labeled with descriptive captions and alt-text in order to communicate their features and conclusions (e.g., of scientific graphs). However, written text describing an image does not usually allow a blind astronomer to make their own conclusions about an image. This indirectly tells folks who are blind or sight-impaired that they do not belong in astronomy. 

\par Blind astronomers and engineers have been developing sonification programs that allow them to more easily and equitably participate in astronomy research and outreach. In astronomy, Dr. Wanda Diaz Merced was one of the first to create a program that added sound to her data when she started losing her sight in graduate school \citep{diaz2013sound}. These programs allow information to be represented by sound tones instead of by visuals. They have since been expanded by institutions to make sonification programs open access. The use of many sound qualities (i.e., tone, overtone, instrument, volume) allow multiple dimensions of data to be represented at once. While data sonification programs are critical for blind and sight-impaired astronomers, they have an added benefit for all astronomers: the human brain has an easier time picking up patterns hidden in 2D noise when data is represented tonally \citep{diaz2013sound}. We expand our ability to identify trends and patterns when we are not limited to representing data in a traditional way.

\par Similarly, to engage with blind and sight-impaired members of the public, it is important that we do not only use processed images or telescope viewing because these are only accessible to individuals who can see. While one can explain an image, they are most likely going to rely on terms that still require some knowledge of sight (i.e. color), and this doesn’t allow for an individual to explore on their own. Just like with data, images can be sonified. The amount of flux in an image can be represented by the volume of sound. Different elements or kinds of gas can be represented by different tones or instruments. When applying sonification, it is critical that the sound elements added are useful and have meaning so as to not be simply performative.

\par In addition to sound, one can make tactile models, where different properties are represented by different elevation and/or texture. Noreen Grice of the Boston Planetarium is pioneering tactile models in addition to creating astronomy books in Braille \citep{grice20153d}. In addition to making the field more accessible, creating models or data products that rely on different senses also pushes astronomers to be more effective science communicators. Examples of tools that can be used to create data products that do not rely on sight are listed in the Appendix.

\subsubsection{Astronomy}
\par Along the axis of representation, we can also think about how astronomy as a field is represented and the motivations we present to folks about why it is important to study. Often times, the narrative is that astronomy answers the most fundamental questions of the universe. While this is very motivating to some, others perceive that this framing makes astronomy feel disconnected from their daily lives and from their communities. Astronomy has so many connections to culture (mythology, holiday calendars, pop culture) that if we acknowledge these connections, we can help folks see how their daily lives and cultural beliefs are connected to the Earth’s place and movement in the Universe. This is another clear opportunity to bring in intersectionality: instead of telling folks that their cultures and identities need to be separate from a science, we can value these connections as a part of astronomy. 

\par Astrology is an interesting case study in how we can bring people into the field instead of isolating them while maintaining the distinction between science and belief. Astrology is not a science. However, it has clear connections to astronomy. It highlights how humans throughout history and all over the world have studied how the sky changes throughout the year. It carefully tracks the motions of the Earth and planets across background stars. This aspect is astronomy. Instead of stating that people cannot believe astrology and be a scientist at the same time, we can discuss how their beliefs intersect with astronomy and where astrology deviates from being a science.

\par This might not seem relevant to disability, but in recognizing the human aspects of astronomy, we also recognize that we, the scientists, are humans with bodies and minds that have needs that cannot always be separated from our work. This recognition helps increase the inclusion of disability. 

\subsection{Engagement}
\subsubsection{Astronomers and Astronomy}

\par Along the axis of engagement, we can first think about how astronomers interact with astronomy. The majority of the time, this involves being in a physical space, whether it is an academic or research institution, a lab, or an observatory. In the United States, there is an assumption that the American with Disabilities Act (ADA) ensures that all buildings are physically accessible, so one doesn’t need to worry about physical accessibility. Unfortunately, this assumption is not correct: while compliance with the ADA offers some bare-minimum accommodations, full accessibility for all individuals frequently requires going beyond the ADA's minimum provisions. Many laboratories, observatories, and academic institutions were built before the ADA; even if barriers have been removed in a given space, as stipulated by the ADA, it does not mean the space is easily accessible. \citep{drummy2018challenging, justice}. 

\par Another common assumption is that if people are in the room, the room is accessible. One should think about disability as a spectrum: an individual’s ability changes throughout their lives on short and long timescales. This means someone might be able to be in a room one day but not another day. Additionally, because of stigma towards disability, an individual might push themselves to be present even though it is bad for their health. 

\par Changes to physical space often take time and require one to navigate the bureaucracy of the institution. During the process of a physical space's negotiation and retrofitting, disabled people are being left out of fully participating in the space. Therefore, whenever possible, changes should be led at the level of an individual department (even if they are only temporary fixes while a more permanent solution is in the works) instead of relying wholly on the institution. This might include changing spaces or traditions or encouraging hybrid participation. By implementing UD and creating options, an individual can choose how they can best participate given the constraints of the physical space. It is very important that all options are seen as equally valid and not just as accessible alternatives. 

\par Giving folks the option and flexibility to participate as needed helps support all individuals, not just disabled ones. Hybrid participation allows someone with childcare or caregiving responsibilities to participate, and people with long commutes can save on commuting time, which is better for the environment and their health. It also encourages people who are sick to stay home and keep everyone healthier.

\subsubsection{Astronomers and Astronomers}
\par Astronomers do not just interact with physical spaces; they also interact with each other. A norm in this field is that there is a heavy reliance on networking, whether it be between talks at conferences, during nights or evenings after conferences, or on social media. This networking and social extroversion requires skills that are not taught. When skills are not taught, they are assumed. This is particularly harmful for Autistic and neurodivergent folks, for whom these unwritten social rules can be particularly opaque, and for physically disabled or chronically ill folks who might not have the energy for social interactions after a busy day. While social networking is not possible for everyone, it is an important part of how someone joins a new project, collaboration, or job. 

\par One way we can make this aspect of astronomy more accessible is by first acknowledging that social networking takes both skills and energy. We can hold workshops or other teaching opportunities for folks to learn and practice these skills. In addition to giving support for these skills, the field should change seeing this networking as a norm. While for many it is fun to go out to karaoke after a day at a conference, for others, this is not possible (beyond disability this can include, folks who are caregivers, folks who don’t drink or don’t feel comfortable being around coworkers and more senior people while they are drinking). We can normalize saying no to social events and ensure that folks who don’t participate in social networking do not miss out on opportunities. We can promote other ways of self promotion, like blog posts or informal emails.  

\subsubsection{Astronomy and the Public}
\par We can also examine how the public engages with astronomy. The current norm is that the public sees the very beginning of research (looking through a telescope) or its finished projects (planetarium shows, processed images, popular science). As mentioned earlier, these both rely on sight, and additionally, observatories are often the least accessible buildings. Smaller telescope viewing events are often put in places away from infrastructure, and therefore inaccessible spaces, to be away from light pollution. 

\par Furthermore, by limiting the public to the initial and end results of research, they don’t get access to the process of research itself. This process, while it can be slow and even boring, is important because it reveals the thought process of researchers: how one formulates an experiment and inevitably how one deviates and adjusts. Seeing the process of research helps the public understand why projects are often delayed and allows people potentially interested in astronomy to evaluate if the field is accessible for them. For example, if one’s only experience with astronomy is not being able to enter an observatory because of stairs, that person might think they cannot do astronomy. Similarly, if a student has only been exposed to the triumphant final products of research, they might question whether they belong in the field when they find themselves struggling in class or during their own research. Creating multiple pathways to join research outreach events (like via Zoom) allows an individual to participate regardless of the accessibility of where the research is taking place and at the same time demystifies science. We should also create events that reveal the beginning and end stages of research to the public.

\par Lastly, the astronomers who are encouraged by their mentors or by the field to participate in outreach tend to be neurotypical, socially outgoing, and not visibly disabled. The norm is that we want folks who will be deemed by society as "cool" to represent the field in order to increase popularity in astronomy. This feeds disability stigma and results in the majority of science social media being done by non-disabled individuals. When the public doesn’t see themselves represented, we are indirectly telling them they don’t belong or won’t be given a voice in astronomy. While no one should be forced to participate in social media or outreach, the field should be more inclusive about who is encouraged to do so. 

\subsection{Action and Expression}
\subsubsection{Evaluation of Success}
\par Along this axis, we can evaluate norms regarding how astronomers show success. The biggest norm here is judging someone’s success by the number of papers one has published. This creates grind culture, in which one is encouraged to be as productive as possible and place productivity over their health, non-work life, and other obligations. Grind culture is unhealthy for everyone, but it is not possible for disabled people who cannot ignore the needs of their bodies and minds. When disabled people honor their health needs, they inevitably have less time to be hyper-productive. This often results in disabled people being seen as lazy or not hardworking, which is not true. Placing productivity as the top priority is also not possible for caregivers or people with community obligations. Thus, the norm of evaluating one’s success based on the number of papers they have is actually evaluating who grind culture benefits most, which is seen as a tool of white supremacy \citep{benjamin2023black}.

\par A way to implement UD here is to genuinely expand how folks are evaluated and dismantle grind culture. This can include looking at the community service one has done for the field including outreach and software development, honoring conference papers and proceedings more, acknowledging that there are different length research projects, honoring research notes showing null results, and formally acknowledging an individual’s personal circumstances. Not only does expanding the way we quantitatively measure success make the field more inclusive to disabled folks, but it will also better honor the work folks do that doesn’t result in papers. The Astro2020 Panel on State of the Profession and Societal Impacts offers a definition of success as "the equitable optimization of knowledge, infrastructure, and innovations, and includes technical and nontechnical contributors and stakeholders, which produce higher quality and more innovative outcomes," \citep{decadel} which can be a guiding definition for creating multiple pathways through which one can show their success in astronomy. 

\subsubsection{Career Advancement}
\par Another norm that exists in the field of astronomy is that at each career stage, one is expected to move institutions. Changing institutions often requires one to move to a different state or even a different country. Moves are stressful, expensive, and physically and emotionally taxing for the majority of people. That stress and exertion is particularly harmful to disabled folks. Additionally, every time a disabled astronomer changes an institution, their healthcare changes. This change means they have to navigate a new insurance network and healthcare system to ensure their care is still covered. It can mean they also need to find new providers (often specialists with long wait times), which can result in a dangerous lapse in medication or care. Each time a disabled person moves to a new institution, they have to navigate the accessibility and support of the department and institution and submit paperwork to HR if they have a formal accommodation. A disabled astronomer needs to worry about the accessibility of the homes and the city/town that their new position is in as well. Accessible homes are often more expensive and limited in number. Disabled astronomers leave behind communities, caregivers, and familiar routines that are an important part of their support plans and well-being. Having to move up to four times in two decades places an extreme burden on a disabled individual.

\par Moving, of course, places a greater burden on other marginalized astronomers as well (i.e., those with dependents and individuals who risk facing higher levels of racism and/or homophobia and less gender-affirming care based on the state they move to). By implementing UD and expanding pathways to career advancement, astronomy can become more inclusive of these folks. We can normalize remote positions, longer postdoctoral positions that replace shorter multiple positions, and staying at one institution for multiple career stages. 

\section{Discussion}

\begin{table*}[t]
    \centering
    \renewcommand{\arraystretch}{1.3} % Adjusts row height for better readability
    \begin{tabular}{|p{5cm}|p{5cm}|p{5cm}|}  % Adjust column widths to fit the page
        \hline
        Representation and Motivation & Engagement & Action and Expression \\
        \hline
      Who can do astronomy?  & Where are events located? & How do we define success? \\
       How do we convey data? & Is networking and socializing expected? & How do we support work life balance? \\
        Why should people care about astronomy? & Does outreach accurately convey the process of research? & How often are folks expected to move institutions? \\
        \hline
    \end{tabular}
    \caption{A table highlighting a summary of Section \ref{sec:pathways}. To make Astronomy more accessible, the field should create multiple pathways along representation, engagement, and action and expression.}
    \label{tab:summary}
\end{table*}

% \begin{figure}
%     \centering
%     \includegraphics[width=1\linewidth]{images/summary.pdf}
%     \caption{Diagram showing summary of Section \ref{sec:pathways}. To make Astronomy more accessible, the field should create multiple pathways along representation, engagement, and action and expression.}
%     \label{fig:summary}
% \end{figure}

A fundamental component of the UD mindset is iteration. Disability, while common, is so nuanced that you cannot foresee every accessibility need or perfectly plan for those you do foresee. Having someone request an accommodation or highlight a barrier they are facing is not a failure; it's part of the process. What's most important is to implement UD whenever possible and to create an environment in which disabled people feel supported and comfortable giving feedback. To create such an environment, it is important to be flexible with your policies. This flexibility indicates that change is possible and that the department values individual circumstances over tradition or policies. Be sure not to reinforce stigma. it's best to mirror someone's language when describing them or their disability. Make it clear that the department will work to make sure everyone can succeed regardless of ability. Ask for, accept, and incorporate feedback often and in meaningful ways.   

\par Many of the ideas presented in this work suggest a larger restructuring of academia; "collective accountability is not just desirable, but necessary if we want academic life to change for the better" \citep{price}. While large changes are needed in order to create a field in which disabled people have the same chance at success as non-disabled folks, it's important to remember that small and local changes are vital as well. The structural changes will take time and persistent effort from many in the field, but the changes that can be made at the department level can be implemented much faster. It is important to prioritize both kinds of changes. 

\appendix \label{sec:appendex}
\section{Starting Questions to Evaluate for Accessibility}

With this section, we aim to create a list of questions a department or institution can use to begin to evaluate itself for accessibility. This is not an exhaustive list, and it is critical that each member of the department is asked about their needs and about any barriers they might have or are currently encountering. Whenever an answer to a below question indicates that a change needs to be made, it is important that the change is permanent and not temporary. Additionally, when making a change, it is a great opportunity to implement UD and create multiple pathways or options for folks. 

\subsection{Institutions}
\begin{itemize}
    \item Are talks/events held in accessible places? Are these places close to office space?
    \item Are lab spaces and telescope rooms actually accessible?
    \item If events are held off campus, are they checked for accessibility (parking, stairs, distance from subway/bus, noise level, etc)?
    \item For activities that usually require changing buildings, is there an accessible alternative that is clear, well known, and supported? Are folks given enough time to change spaces?
    \item Are department talks streamed/recorded? 
    \item Are speakers given a microphone?
    \item When preparing for talks or making signs for the department, are there checks for colorblindness?
    \item Are folks encouraged or required to stay home while sick without facing any penalties?
    \item Does the department place work over one’s physical health?
    \item Does your department have a relationship with your campus’s Disability Resource Center/HR? Do you remind (grad) students + faculty that they have access to it?
    \item Are visitors asked about their accessibility needs (speakers, job applicants, prospective grads) prior to their arrival?
    \item When talking about equity, inclusion, and diversity, is the axis of disability included?
    \item Are ableist sayings or co-opted disability identity language a part of the department vernacular? 
    \item When an accessibility concern is brought to your attention, is the change that's made temporary or permanent?
\end{itemize}

\subsection{Conferences}
\begin{itemize}
    \item Can people participate remotely?
    \item Are participants asked for access needs and do you take responsibility for fulfilling them?
    \item Are talks streamed/recorded? 
    \item Are speakers given a microphone? 
    \item Are the length of timed talks flexible for folks with speech impairments?
    \item Are sign language interpreters or transcribers provided?
    \item Are there equally valid accessible break activities?
    \item Are people expected to socialize during and after the conference events?
    \item Are folks encouraged to move as they need? (Ex: knit during talks/meetings)
    \item Are folks given enough time to changes spaces?
    \item Are there substantial and frequent breaks?
    \item Are there quiet rooms?
    \item Are masks provided or required?
\end{itemize}

\subsection{Outreach }
\begin{itemize}
    \item Are outreach events held in physically accessible places?
    \item Does the event rely on one sense? Is information presented in multiple forms?
    \item Are speakers given a microphone?
    \item Does the event require participants to stand for long periods of time or navigate in the dark?
\end{itemize}

\section{Tools for Increasing Accessibility}

There are many tools available that can increase the accessibility of products and events, from documents to websites to models. Below, we highlight some examples that are especially relevant to the field of astronomy and UD. There are many others not listed below. 
\begin{itemize}
    \item Sonification: General information and examples of projects: \url{https://sonificationworldchat.org/}. Some of the Python programs and apps can be found here \url{https://www.audiouniverse.org/research/other-sonification-projects}. They range from sonfiying spectra to sonifying data points. An example of this kind of sonification can be found here: \url{https://chandra.si.edu/sound/casa.html}.
    \item Tactile models: Some free 3D printer files of astronomical tactile models can be found here: \url{https://tactileuniverse.org/models/}.
    \item Colorblindness: There are many tools that can either simulate colorblindness for the user to then evaluate their own color scheme or automatically evaluate a color scheme. These tools include Color Oracle (\url{https://colororacle.org/}).
    \item American Sign Language accessibility: utilize Deaf-led dictionaries and resources such as Atomic Hands (\url{https://atomichands.com/}); see further recommendations in \citet{Rice2024Including} and \citet{Sfarnas}.
    \item Websites: Browser extensions or app tools like WAVE Web Accessibility Evaluation Tools (\url{https://wave.webaim.org/}) can, among other things, help evaluate how effectively a screenreader can access information on a website.
    \item Research outreach: Shadow the Scientists directly connects that public and scientists via Zoom as research is being conducted. This allows disabled folks to participate how and where is more accessible to them while creating transparency in how research happens. \url{https://shadow.ucsc.edu/}
\end{itemize}

\section{Ableist Phrases and Language to Avoid}

It is important to not use the below phrases when talking about disabled people or when joking about non-disabled people. This list is not exhaustive. Whenever talking about disability, it is important to use language that aligns with how an individual describes themselves. Language is personal, and each individual will have preferred terms \citep{brown}. 

\begin{itemize}
    \item blindspot or similar: what folks mean is a gap in understanding, not the inability to see
    \item handicapped: this word is outdated and shouldn't be used to refer to people, parking spaces, or areas
    \item crippled: many in the disability community have reclaimed this word, but it is not for non-disabled people to use
    \item differently-abled or special needs: the disability community prefers the term "disabled"  
    \item insane or crazy: the use of these terms to describe someone's behavior add to the stigma around mental illness
    \item deaf ears: deaf is not an ableist word when describing someone with a hearing disability. However, it is ableist to use phrases like "deaf ears" because you mean not listening or ignorant, not unable to hear.
    \item low/high functioning: this is an outdated and inaccurate way to categorize Autistic people
    \item suffering from: it is ableist to assume that someone with a disability is inherently suffering and therefore disability is inherently bad
    \item tone deaf: this phrase links deafness with ignorance; instead say ignorant or similar
    \item wheelchair bound: this phrase is inaccurate; instead say wheelchair user
    \item stutter: while the word "stutter" is not offensive, phrases like "did I stutter" equate stuttering to hesitation or a lack of confidence, which is not the cause of a stutter
\end{itemize}

\bibliography{disability}{} 
\bibliographystyle{aasjournal}

\end{document}